\documentclass[aps,prd,floats,floatfix,showpacs,superscriptaddress,nofootinbib]{revtex4}
\usepackage{amsmath,amsfonts}
\newtheorem{theorem}{Theorem}
\newtheorem{lemma}{Lemma}

\begin{document}

\title{ With commuting Killing vectors,  the lapse and shift of one Killing vector are constants along the other.}

\author{Niall \'{O} Murchadha}{\footnote {niall@ucc.ie}}

\affiliation{Physics Department, University College, Cork, Ireland}

\begin{abstract}
Given an d-dimensional manifold with two commuting Killing vectors, together with an d - 1 dimensional submanifold in which one of the Killing vectors lies,  then the lapse and shift of the second Killing vector, relative to this slice,  remain constant along the orbits of the `surface' Killing vector. Alternatively, the six dot products that can be formed from the three vectors, the two Killing vectors and the normal to the submanifold, are all constants along the `surface' Killing vector.
\end{abstract}

\maketitle
\section{The Calculation}
Consider the Kerr solution of General Relativity. This has two Killing vectors which commute. One is axially symmetric, the other can be chosen to be timelike, at least near infinity. These are not uniquely specifiable, any linear combination of Killing vectors is another pair of commuting Killing vectors. Nevertheless, it is easy to find coordinates, $(t, r, \theta, \phi)$, in which the timelike Killing vector, $T^{\mu}$, is $T^{\mu} = (1, 0, 0, 0)$ and the axial Killing vector, $R^{\mu}$, is $R^{\mu} = (0, 0, 0, 1)$, and the metric components only depend on $r$ and $\theta$.  The normal to the $t = 0$ slice, $n^{\mu}$ satisfies $n_{\mu} \propto (1, 0, 0, 0)$. This means that the axial Killing vector lies in the slice, i.e., $n_{\mu}R^{\mu} = 0$.  Further, the lapse, $\alpha, g^{00} = -\alpha^{-2}$, and shift, $\beta, \beta_i = g_{0i}$, of the 4-metric are the projections of the time-translational vector, $t^{\mu} = (1, 0, 0, 0)$ relative to the slice, $t^{\mu} = \alpha n^{\mu} + \beta^{\mu}$ with $n^{\mu}\beta_{\mu} = 0$. Of course $t^{\mu} = T^{\mu}$, the `time' Killing vector. The metric, the lapse and shift, and thus the projections of the Killing vector, $T^{\mu}$,  depend only on $r$ and $\theta$. Therefore the $\phi$ derivative of both the lapse and shift vanish. In other words, the projections of the $T^{\mu}$ Killing vector are constants along the orbits of the $R^{\mu}$ Killing vector.

I wish to show that this is true it general. It does not depend on the dimensionality, on the signature, on the geometry, or on the topology of the manifold, and the Einstein equations play no role. 

This result emerged as a small part of a calculation of the behaviour of `1 + log' slices in spacetimes with a helical Killing vector \cite{B}. I felt that it had more general interest, so I decided to write it up independently. After the article was finished, I discovered that essentially the same result was obtained by Robert Beig and Piotr Chru\'sciel ten years ago \cite{beig}.

The result is as follows: Given an d-dimensional manifold, $M$, which contains  two Killing vectors. I will assume that the signature of the manifold is $(-, +, + , \dots)$, but the result does not depend on this. One Killing vector, denoted by $R^\mu$, I assume to be spacelike, while the other, denoted by $T^\mu$, can be timelike, spacelike, or fluctuate. I assume that the Killing vectors commute, see e.g. (\cite{W} Eq.(7.1.2)). I further assume that  $R^{\mu}$ lies in some spacelike d - 1 dimensional submanifold, $S$. This has a timelike unit normal $n^{\mu}$, such that $n_{\mu}R^{\mu} = 0$. One can decompose the $T^{\mu}$ Killing vector relative to $n^{\mu}$ as $T^{\mu} = \alpha_Tn^{\mu} + \beta_T^{\mu}$, where $\beta_T^{\mu}n_{\mu} = 0$. These are the lapse and shift of the d-dimensional metric generated by dragging the d - 1 submanifold along the $T$ Killing vector. I will show that the lapse of the $T$ Killing vector, $\alpha_T = n_{\mu}T^{\mu}$,  the magnitude of the shift, $\beta_T$, and the projection of the shift, $\beta_{\mu}^TR^{\mu}$, are all constants along the orbits of the  $R$ Killing vector.

Before I reach the final conclusion, I need two intermediate results:
\begin{lemma}
If I have a manifold with a Killing vector $R^{\mu}$ and a  vector field, $v^{\mu}$,  which is perpendicular to the Killing vector, then the `acceleration' of the curve congruence defined by $v^{\mu}$, $a^{\mu} = v^{\mu}_{;\alpha}v^{\alpha}$, is orthogonal to the Killing vector.
\end{lemma}
This is a one-line calculation:
\begin{equation}
a^{\mu}R_{\mu} = v^{\mu}_{;\alpha}v^{\alpha}R_{\mu} = (v^{\mu}R_{\mu})_{;\alpha}v^{\alpha} - v^{\mu}v^{\alpha}R_{\mu;\alpha} = 0.
\end{equation}
The first term in the last expression vanishes because $v^{\mu}R_{\mu} = 0$ and the second term vanishes because $v^{\mu}v^{\alpha}$ is symmetric in $(\mu, \alpha)$ while $R_{\mu;\alpha}$ is antisymmetric due to the Killing equation,  $R_{\mu;\alpha} + R_{\alpha;\mu} = 0$ .

Note that I do not assume that $v^{\alpha}$ is surface forming. However, if it is I can prove something further:
\begin{lemma}
If I have an d-dimensional manifold with a Killing vector $R^{\mu}$, and an d-1 dimensional submanifold defined by a unit  timelike vector field, $n^{\mu}$, which is orthogonal to the Killing vector, so that $n_{\mu}R^{\mu} = 0$, then $R^{\delta}(n_{\gamma;\delta} - n_{\delta;\gamma}) = 0$.
\end{lemma}
This is a straightforward application of the Frobenius Theorem, see e.g. (\cite{W} Appendix B.3). Since $n^{\mu}$ is surface-forming, the `twist' of it must vanish. The twist is the doubly projected part of $(n_{\gamma;\delta} - n_{\delta;\gamma})$. This is 
\begin{eqnarray}
0 = (n_{\alpha;\beta} - n_{\beta;\alpha})^{PP} & = &(n_{\gamma;\delta} - n_{\delta;\gamma})(\delta^{\gamma}_{\alpha} + n^{\gamma}n_{\alpha})(\delta^{\delta}_{\beta} + n^{\delta}n_{\beta}) \nonumber \\
& = &n_{\alpha;\beta} - n_{\beta;\alpha} + a_{\alpha}n_{\beta} - a_{\beta}n_{\alpha}\label{frob}
\end{eqnarray}
(An easy check of this expression is to multiply by either $n^{\alpha}$ or $n^{\beta}$ and show that it vanishes, using that $a^{\alpha} = n^{\alpha}_{;\beta}n^{\beta}$).
Note the last two terms are orthogonal to $R^{\alpha}$, the first because of lemma 1, the second because $n_{\alpha}R^{\alpha} = 0$ so one gets that $R^{\delta}(n_{\gamma;\delta} - n_{\delta;\gamma}) = 0$ if one has an n-1 surface which contains the Killing vector. 

(Aside: It is easy to see that the doubly projected part of the curl vanishes if we assume that the surface is the level set of some scalar $V$. We then have $n_{\mu} = f\nabla_{\mu} V$, where $f$ is related to the norm of $\nabla V$. Substitute this into Eq.(\ref{frob}) to get $(n_{\alpha;\beta} - n_{\beta;\alpha})^{PP} = (f[V_{;\alpha\beta} - V_{;\beta\alpha}] + f_{;\alpha}V_{;\beta} - f_{:\beta}V_{;\alpha})^{PP}$. The first two terms cancel because two covariant derivatives on scalars commute, and the other two terms vanish because $V_{;\beta}$ is parallel to $n_{\beta}$ and thus gets killed by one of the projection operators, while the other term gets killed by the other projection.)

The key result is as follows:
\begin{theorem}
Given an d-manifold with two commuting Killing vectors $R$ and $T$ and an  d-1 submanifold through it with $n{^\mu}$ as its unit normal such that $R^{\mu}n_{\mu} = 0$. Then (a) the lapse of the $T$ Killing  relative to this slice, $\alpha_T = n_{\mu}T^{\mu}$, satisfies $R^{\mu} \nabla_{\mu} \alpha_T = 0$, (b) the Killing shift satisfies $R^{\mu}\nabla_{\mu}(\beta_T^{\nu}\beta^T_{\nu}) = 0$, and (c) $R^{\mu}\nabla_{\mu}(\beta_T^{\nu}R_{\nu}) = 0$.
\end{theorem}
(a) I start with the commutator of the Killing vectors, which I know vanishes, take the dot product with the normal to the surface and manipulate slightly.
\begin{equation}
0 = n_{\gamma}(R^{\delta}\nabla_{\delta} T^{\gamma} - T^{\delta}\nabla_{\delta}R^{\gamma}) = R^{\delta}\nabla_{\delta}(n_{\gamma}T^{\gamma}) - T^{\delta}\nabla_{\delta}(n_{\gamma}R^{\gamma}) - R^{\delta}T^{\gamma}(n_{\gamma;\delta} - n_{\delta;\gamma})
\end{equation}
This gives the desired result. 
The first term equals $R^{\delta}\nabla_{\delta} \alpha_T $; the second term vanishes because $n_{\gamma}R^{\gamma} = 0$  and the third and fourth terms vanish because of Lemma 2.

(b) To prove the first shift result, I start by showing that the length of the $T$ Killing vector is a constant along the orbits of the $R$ Killing vector. We have, using the fact that the Killing vectors commute
\begin{equation}
R^{\mu}\nabla_{\mu}(T^{\nu}T_{\nu}) = 2 R^{\mu}T^{\nu}\nabla_{\mu}T_{\nu} = 2T^{\mu}T^{\nu}\nabla_{\mu}R_{\nu}
\end{equation}
However, this expression is zero because $T^{\mu}T^{\nu}$ is symmetric in $(\mu, \nu)$ while $\nabla_{\mu}R_{\nu}$ is antisymmetric because $R$ is a Killing vector.
We know that $T^{\mu}T_{\mu} = \beta^2 - \alpha^2$. We have that both $T^2$ and $\alpha$ are constants along $R$.  Therefore $\beta^2$ must also be constant.

(c) To prove the second shift result I start with the realization that the dot product of the $R$ Killing vector with the `T' shift equals the dot product of the two Killing vectors, i.e.,
\begin{equation}
R_{\mu}T^{\mu} = R_{\mu} (\alpha_Tn^{\mu} + \beta_T^{\mu}) = R_{\mu} \beta_T^{\mu}.
\end{equation}
Now
\begin{eqnarray}
R^{\mu}\nabla_{\mu}(R^{\nu}T_{\nu})& = &R^{\mu}R^{\nu}\nabla_{\mu}T_{\nu} +  R^{\mu}T^{\nu}\nabla_{\mu}R_{\nu} = R^{\mu}R^{\nu}\nabla_{\mu}T_{\nu} -  R^{\mu}T^{\nu}\nabla_{\nu}R_{\mu}\nonumber \\
& = &R^{\mu}R^{\nu}\nabla_{\mu}T_{\nu} -  R^{\mu}R^{\nu}\nabla_{\nu}T_{\mu} = 0.
\end{eqnarray}
I first use the Killing equation on $R$ to switch indices, and then use the fact that $R$ and $T$ commute to switch them. I finally get the zero because we have a symmetric term multiplying an antisymmetric term. 

There is one other object which is constant. This is the length of the $R^{\mu}$ Killing vector. It is easy to see
\begin{equation}
R^{\nu}(R^{\mu}R_{\mu})_{;\nu} = 2 R^{\nu}R^{\mu}R_{\mu;\nu} = 0.
\end{equation}
This holds for the same reasons; $R^{\nu}R^{\mu}$ is symmetric in $\nu$ and $\mu$ while $R_{\mu;\nu}$ is antisymmetric in $\nu$ and $\mu$ because of the Killing equation.

This is the best that one can do. I cannot hope to prove anything about the components of $\beta$ without introducing coordinates. We know that $\beta^{\mu}$ is a vector, and therefore the components depend on the choice of coordinates. If the coordinates in the surface do not respect the $R^{\mu}$ symmetry, neither will the components of $\beta^{\mu}$. The choice of $T$ and $R$ is somewhat arbitrary. If one adds any multiple of $R$ to $T$, one still gets commuting Killing vectors and the theorem continues to hold. In this case $\alpha_T$ does not change, only $\beta_T$. Further, if one has one submanifold in which $R$ lies, one has a whole family of them. This will change $n^{\mu}$, and thus $\alpha_T$ and $\beta_T$. The new $\alpha_T$ and $\beta_T$ are still constants.

There are three special vectors in the manifold, $T^{\mu}, R^{\mu}$, and $n^{\mu}$. We have six dot products. An alternative statement of the result proven here is that each of these six dot products are constants along the $R^{\mu}$ Killing vector. Two of the dot products are trivial because $R_{\mu}n^{\mu} = 0$ and $n_{\mu}n^{\mu} = -1$. I have dealt with $n^{\mu}T_{\mu} = \alpha_T$ in Part (a) of the Theorem, with $T^{\mu}T_{\mu}$ in Part (b) of the Theorem, and with $R^{\mu}T_{\mu} = R^{\mu}\beta_{\mu}$ in Part (c) of the Theorem. I prove the constancy of $R^{\mu}R_{\mu}$ just after the Theorem.

When one numerically evolves initial data for the gravitational field so as to generate a solution of the Einstein equations, one has to specify the lapse and shift. A popular choice to evolve the lapse is the so-called `1 + log' slicing \cite{Bo}. This is of the form
\begin{equation}
\mathcal{L}_{n^{\mu}} \alpha = bK
\end{equation}
where $\mathcal{L}$ is the Lie derivative, $n^{\mu}$ is the unit timelike normal to the slice, $b$ is some constant (2 is a standard choice), and $K$ is the trace of the extrinsic curvature. Choosing any `time' vector, this can be written as
\begin{equation}
(\partial_t - \beta^i\partial_i)\alpha = b\alpha K
\end{equation}
where $\beta^{\mu}$ (expressed in preferred coordinates) and $\alpha$ are the lapse and shift of the `time' vector $t^{\mu} = (1, 0. 0 \dots)$ (again in preferred coordinates). If the manifold has a Killing vector, $T^{\mu}$, a `stationary 1 + log' slice will be one where, if I use $T$ as my `time' vector, I have that $\partial_T \alpha = 0$ in the 1 + log gauge. In other words, a `stationary 1 + log' slice is one which satisfies
\begin{equation}
K = -\frac{\beta^i_T\partial_i \alpha_T}{b\alpha_T} \label{1 + log}
\end{equation}

Consider a stationary axisymmetric spacetime and assume we find an axisymmetric `stationary 1 + log' slice, i.e., one which satisfies Eq.(\ref{1 + log}). The `time' Killing vector, $T^{\mu}$, has some arbitrariness. We can always add any multiple of the `rotation' Killing vector to it, i.e., $T'^{\mu} = T^{\mu} + cR^{\mu}$, where $c$ is any constant. This given slice is `1 + log stationary' with respect to the new as well as to the old `time' Killing vector. This is because $\alpha_{T'} = \alpha_T$ and $\beta^i_{T'} = \beta^i_T + cR^i$. Now we can see
\begin{equation}
-\frac{\beta^i_{T'}\partial_i \alpha_{T'}}{b\alpha_{T'}}  =  -\frac{(\beta^i_T + cR^i)\partial_i \alpha_T}{b\alpha_T}= -\frac{\beta^i_T\partial_i \alpha_T}{b\alpha_T} = -K,
\end{equation}
because the aim of this article was to show $R^i\partial_i \alpha_T = 0$.

\section{Discussion}
Given a spacetime with two commuting Killing vectors, one can choose coordinates so that  the metric depends on only two coordinates, $g_{\mu\nu}(x^2, x^3)$, see e.g. (\cite{W} Eq.(7.1.3)). In this coordinate system the $T$ Killing vector is $T^{\mu} = (1, 0, 0, 0)$ and  $R$  is $R^{\mu} = (0, 0, 0, 1)$. The normal to the $t$ = constant slices satisfies $n_{\mu} \propto (1, 0, 0, 0)$, so obviously $n_{\mu}R^{\mu} = 0$. Since $g_{11}$ and $g_{1i}$ only depend on $(x^2, x^3)$, one finds that $\alpha_T $ (and also $\beta_T$) is independent of $x^4$, and therefore it is constant along the $x^4$ direction. We immediately recover the theorem in this special case.

Given one such coordinate system, one can change the coordinates by $x'^1 = x^1 - h(x^2, x^3), x'^2 = x^2, x'^3 = x^3, x'^4 = x^4$, and discover $g'_{\mu\nu} = g_{\mu\nu}'(x^2, x^3)$. This gives a new time slicing in which the $R$ Killing vector lies and the theorem also holds for this slicing. If one could show that all such time slices can be generated by  appropriate functions $h(x^2, x^3)$, then the theorem would be true in general.

A very plausible argument can be made. Consider the $x^1 = 0$ slice in a coordinate system where the metric depends only on $(x^2, x^3)$. Now consider another axisymmetric 3-slice, $S$. Take the 2-surface, in $x^1 = 0$, given by $(0, x^2, x^3, 0)$. Drag each point of this surface along the $T$ Killing vector until one intersects $S$. This should give a surface in $S$ and a function, $f(x^2, x^3)$, the coordinate distance one moves along the Killing vector from $x^1 = 0$ to $S$. Take this 2-surface in $S$ and drag it along the rotational Killing vector. This should generate the surface $S$. I claim that the surface $S$ is defined by $x^1 = f(x^2, x^3)$.

The logic is as follows:  Take a point $(0, x^2, x^3, 0)$ in the original 2-slice and drag that point a coordinate distance $\Delta x^4$ along the rotational Killing vector to $(0, x^2, x^3, \Delta x^4)$.  Now drag this point a distance $f(x^2, x^3)$ along the translational Killing vector. I claim that this point should be on $S$ and is the point found by first moving $f(x^2, x^3)$ along the time Killing vector to $S$ and then $\Delta x^4$ along the rotational Killing vector.  The two motions should commute because the two Killing vectors do.  Therefore all  surfaces in which the $R$ Killing vector lies can be found by an appropriate choice of height function and the Killing lapse is constant along the rotational Killing vector. 

Possible holes I can see in this argument is that the range of $(x^2, x^3)$ might change from one slice to the other, or that the function $f(x^2, x^3)$ might be unpleasant in some way, maybe loss of differentiability, discontinuities, who knows. Might it be that $\alpha_T$ vanish at a point? Because of this residual unease, I think the argument in Section 1 is cleaner.

The Beig-Chru\'sciel approach is as follows:

They consider a pseudo-Riemannian 4-manifold with a spacelike Cauchy surface through it. They assume they have two Killing vectors and compute the lapse and shift of each. They then calculate the commutators of the two Killing vectors in terms of the lapse and shift. Their Eq.(2.15), written in my notation, reads
\begin{equation}
\{(\alpha_T, \beta^i_T), (\alpha_R, \beta^j_R)\} = (\mathcal{L}_{\beta_T}\alpha_R - \mathcal{L}_{\beta_R}\alpha_T, [\beta_T, \beta_R]^l + \alpha_T\nabla^i\alpha_R - \alpha_R\nabla^l \alpha_T) 
\end{equation}
Where $\{ , \}$ represents the commutator of two 4-vectors, and $[ , ]$ represents the commutator of two 3-vectors. In the case I am dealing with I assume that the 4-commutator vanishes and that $\alpha_R \equiv 0$. Then the Beig-Chru\'sciel formula reduces to
\begin{equation}
0 = ( - \mathcal{L}_{\beta_R}\alpha_T, [\beta_T, \beta_R]^l ) 
\end{equation}
The vanishing of the lapse, $\mathcal{L}_{\beta_R}\alpha_T = 0$, is equivalent to part (a) of Theorem 1, and the vanishing of the commutator of the shifts is gives us part (b) and (c).

If we drop the condition that one of the Killing vectors lie in a surface, we can still prove the following result:
\begin{theorem}
If we are given a manifold with two commuting Killing vectors, the three dot products are constants along each of the Killing vectors.
\end{theorem}

Let me end with a challenge to the reader:, i.e., something I believe to be true but cannot prove. If the rough argument laid out in this Section about constructing preferential coordinates can be made precise, it is clear that $\alpha_T$ is independent of both $t$ as well as $\phi$. This implies that $T^{\mu}\nabla_{\mu} \alpha_T = T^{\mu}\nabla_{\mu}n_{\nu}T^{\nu} = 0$. Can anyone come up with a proof?

\acknowledgments
I would like to thank the authors of \cite{B}, especially Thomas Baumgarte. I would also like to thank Bernd Schmidt, who pointed out to me that the result was much more general than I originally thought. NOM acknowledges the support of SFI grant 07/RFP/PHYF148.

\end{document}